\documentclass[10pt]{article}
\usepackage[margin=3.0cm]{geometry}
\usepackage{amsfonts,amsmath,amsthm,amssymb}
\usepackage{mathpazo} 
\usepackage[affil-sl]{authblk}  
\usepackage{graphicx}
\usepackage{subfig}
\usepackage{color}
\usepackage{url}
\usepackage[T1]{fontenc}

\newcommand{\haf}{\ensuremath{\frac{1}{2}}}  	
\newtheorem{exm}{Example}

\def\figs{Figures} 		

\title{A Note on Disk Drag Dynamics}
\author{Neil J. Gunther}
\affil{\small Performance Dynamics Company, Castro Valley, CA 94552 \authorcr njgunther@perfdynamics.com}
\date{\small \today}

\begin{document}
\maketitle
\thispagestyle{empty}

\begin{abstract}
The electrical power consumed by typical magnetic hard disk drives (HDD) not
only increases linearly with the number of spindles but, more significantly, it
increases as very fast power-laws of speed (RPM) and diameter. Since the
theoretical basis for this relationship is neither well-known nor readily
accessible in the literature, we show how these exponents arise from aerodynamic
disk drag and discuss their import for green storage capacity planning.
\end{abstract}

\section{Introduction}
\begin{figure}[!hb]
  \centering
  \subfloat[Linear axes]{\label{fig:dkplot-lin}
  \includegraphics[scale=0.5]{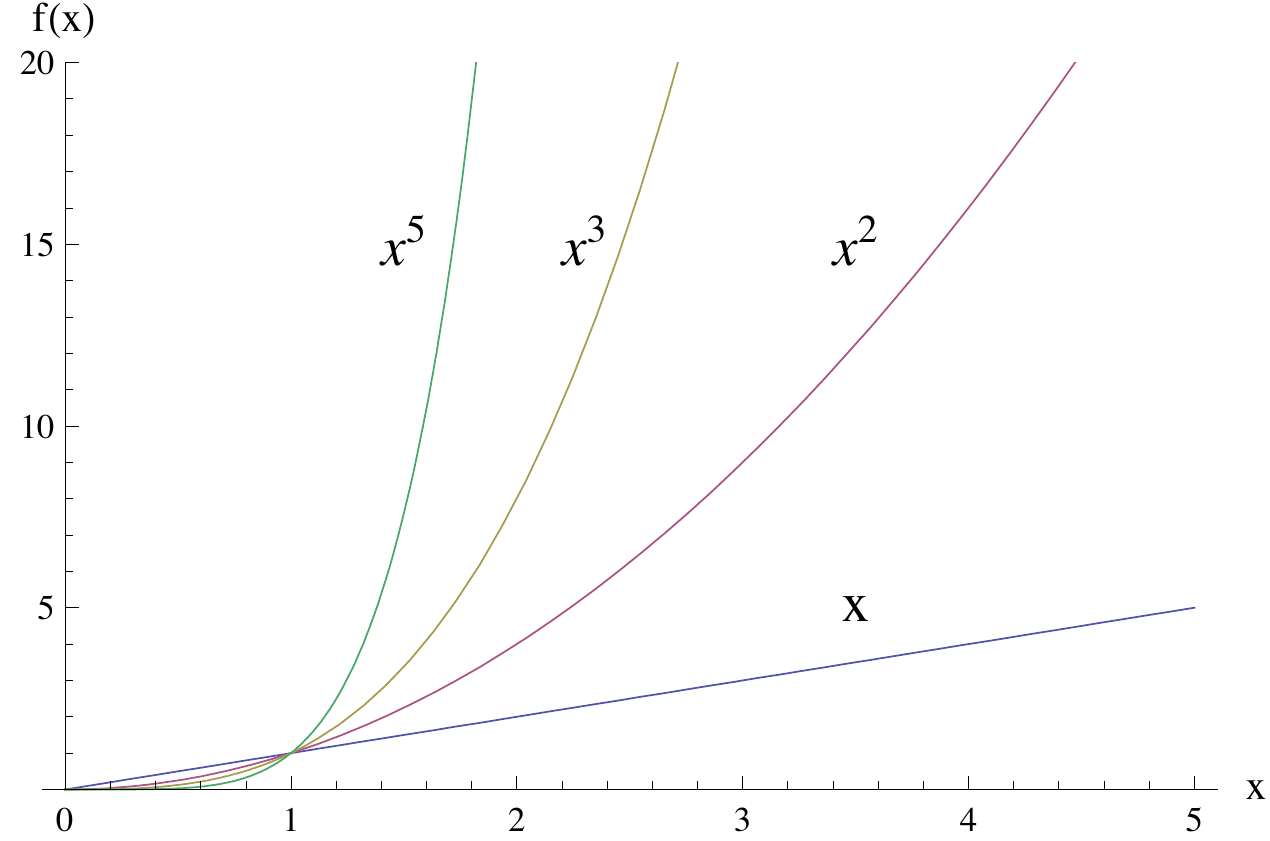}}
  \hspace{1cm}
  \subfloat[Log-log axes]{\label{fig:dkplot-log}
  \includegraphics[scale=0.5]{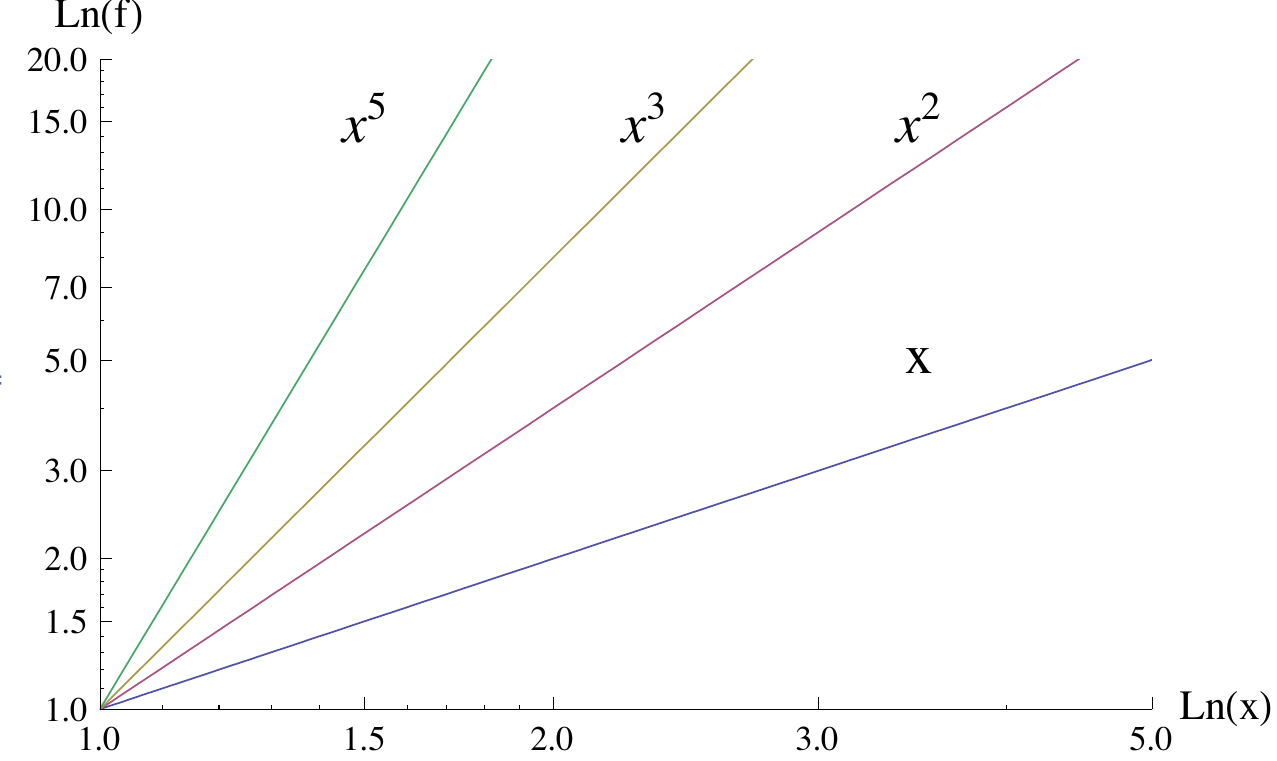}}
  \caption{A fast function family: $f(x)=x^p$ with powers $p=1,2,3,5$.}
  \label{fig:dkplots}
\end{figure}

The semi-empirical relationship~\cite{msc:1988,ieee:1990,ieee:1996}
\begin{equation} 
\text{Power} \propto ~\text{Platters} \times ~\text{RPM}^{2.8} ~\times ~\text{Diameter}^{4.6} \, ,
\label{eqn:emp-model}
\end{equation}
has been used to motivate ``greener''magnetic HDD designs~\cite{phd:2005}.
Of the three variables that can used to match a given
power constraint, diameter and RPM ({\em revolutions per minute}) have the greatest impact due to 
the high degree of their respective positive exponents (Fig.~\ref{fig:dkplots}).
It is noteworthy that \eqref{eqn:emp-model} is 
not a function of capacity (e.g., GB); the typical metric used for storage capacity planning.

The HDDs used in laptops are designed to operate in the energy-limited
environment of a battery-powered system. Consequently, those disks tend to be
smaller, slower, and have fewer platters. Server HDDs, on the other hand, tend
to follow the converse trend. Since the RPM can
be varied dynamically (DRPM)~\cite{phd:2005}, disk drive manufacturers now offer this
capability in some models~\cite{wddisk}.

\section{Calculating Power}
For the purposes of comparing the power consumption of two HDD 
models (say, {\em a} and {\em b})\footnote{Such as the kind of
calculations required for data center and storage capacity planning.} it is more
convenient to express \eqref{eqn:emp-model} in the {\em ratio form}
\begin{equation} 
\dfrac{\mathcal P_{\rm b}}{\mathcal P_{\rm a}} = \bigg( \dfrac{N_{\rm b}}{N_{\rm a}} \bigg)
\bigg( \dfrac{\Omega_{\rm b}}{\Omega_{\rm a}} \bigg)^{2.8} \bigg( \dfrac{D_{\rm b}}{D_{\rm a}} \bigg)^{4.6} \, .
\label{eqn:ratio-model}
\end{equation}
which avoids the necessity of otherwise determining a constant of proportionality.

In this simpler notation, $\mathcal P$ is the power consumed (Watts), $N$ is the number of platters per spindle,
$\Omega$ is the angular speed (RPM), and $D$ is the platter diameter (not the external form factor) usually 
expressed in {\em inches} in U.S. vendor data sheets. Another convenience
of using \eqref{eqn:ratio-model} is that the ratio of two HDD parameters that are common (e.g., 
the same diameter) simplifies to unit value. 

\begin{exm}[RPM Variation] \label{exam:rpm}
The parameters in the following table are for HDDs with 
a single platter $N_{\rm a} = N_{\rm b} = 1$ and the 
same diameter $D_{\rm a} = D_{\rm b} =2.6$ inches~\cite{phd:2005}. 

\begin{center}
{\rm
\begin{tabular}{rrrr|r}
\multicolumn{5}{c}{\rm \bf Single platter 2.6 inch HDD}\\
\hline
$\mathcal P_{\rm a}$  & $\mathcal P_{\rm b}$ & $\Omega_{\rm a}$ & $\Omega_{\rm b}$ & 
\multicolumn{1}{c}{$\mathcal P^*_{\rm b}$} \\
\hline
0.91  & 1.13   & 15,098 & 16,263  &   1.121\\
2     & 35.55  & 19,972 & 55,819  &  35.550\\
35.55 & 499.73 & 55,819 & 143,470 & 499.782\\
\hline
\end{tabular}
}
\end{center}
Hence, only the ratio of the angular speeds contributes in \eqref{eqn:ratio-model} to the estimated power 
$\mathcal P^*_{\rm b}$ in the last column.
\end{exm}

Because the exponents in \eqref{eqn:emp-model} are relatively large, 
the functions of RPM and diameter are highly nonlinear and that means:
\begin{itemize}
\item 
Potentially large energy savings can be achieved even within a limited selection of HDD models,  
especially when taken in aggregate across SANs, NAS, JBoDs, or RAID storage configurations.
\item 
The key disk parameters are available from vendor data sheets, although some caveats still 
apply~\cite{specs}.
\end{itemize}

\begin{exm}[Diameter Variation]
The data used in this example are taken from Tables 3.3--3.5 in~\ref{exam:rpm}. 
The effect of reducing HDD diameter ($d$) is shown in Fig.~\ref{fig:dkplot3D}.
\begin{figure}[!ht]
\begin{center}
\includegraphics[scale = 0.75]{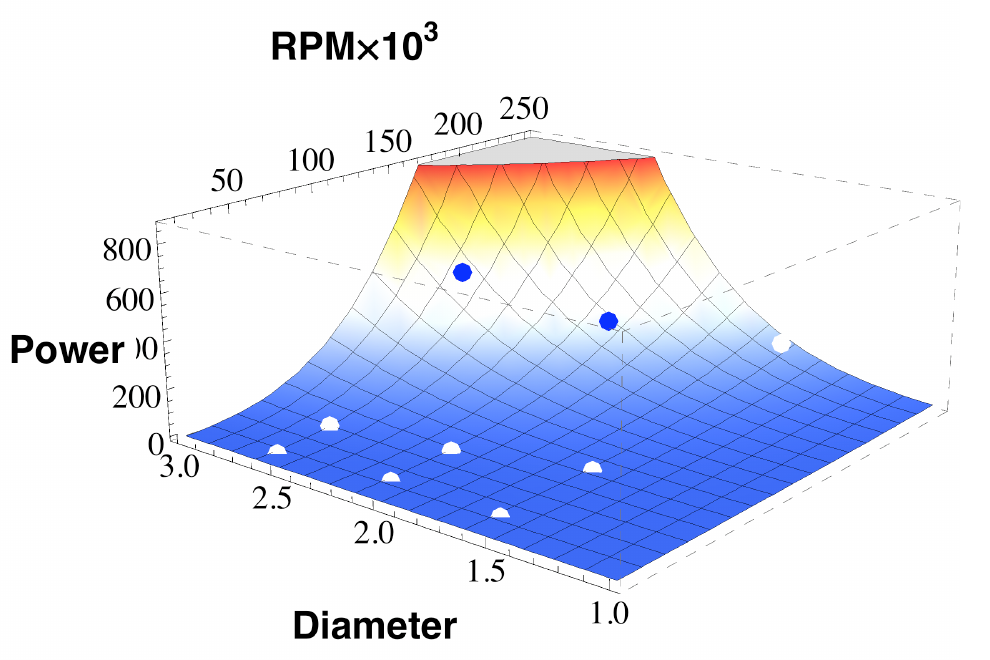} 
\caption{Specific HDD configurations ({\em dots}) plotted on the power surface defined by 
\eqref{eqn:emp-model}. The design trend is to move toward the bottom corner of the HDD power surface.}
\label{fig:dkplot3D}
\end{center}
\end{figure}
At highest RPM, both the $d=2.6$ inch and $d=2.1$ inch
HDDs are located higher on the power hill than the $d=1.6$ inch HDD.
\end{exm}

\section{Theoretical Justification}
We now turn to providing a theoretical justification 
for the semi-empirical relationship in \eqref{eqn:emp-model}.
As a starting point, we assume:
\begin{enumerate}
\item The empirical exponents can be associated with integers (rounded up)
\item Planer rotation of the platter imposes axial symmetry 	
\item Rotational friction is present since the platter is not {\em in vacuo}
\end{enumerate}

A thin rotating platter implies that inertial linear relationships, like the
kinetic energy alone, cannot produce \eqref{eqn:emp-model}. Rather, rotational
quantities, like moments of inertia, are more important.

Rotation also implies that there is aerodynamic friction due to
the platter spinning. This is also assumed to be greater than friction in the spindle bearings. 
Since the HDD platter resides in a stationary
housing, we assume there is no translational drag proportional the cross-sectional
area of the platter; as there would be for a fan-blade or propeller pulling air.

\subsection{Pressure}
The \href{http://www.princeton.edu/~asmits/Bicycle_web/Bernoulli.html}{
Bernoulli equation} tells us that the pressure at some point in a fluid 
(e.g., air with density $\rho$) is given by:
\begin{equation*}
P_{\rho} = p + \rho g h + \haf \rho v^2 \, .
\end{equation*}
For the spinning disk, we are only interested in the dynamic contribution of the 
third term. The external and hydrostatic contributions can be ignored.

Since pressure is force per unit area ($F/A$), 
the aerodynamic frictional force will be 
\begin{equation}
F_{\rho} = \haf \rho v^2 C_d A  \, , \label{eqn:drag}
\end{equation}
where $C_d$, the {\em drag coefficient} is an additional fudge factor that
covers a multitude of sins as to how the fluid drag actually
occurs\footnote{Factors such as: shape, roughness, viscosity, compressibility,
boundary-layer separation, etc.}.

The drag force in \eqref{eqn:drag} is to be understood as being tangential,
rather than centripetal. The velocity $v$ at any point on the platter surface 
has magnitude relative to the air in the tangential direction of the disk spin.
Similarly, the area $A$ is on the surface of the platter (not its edge) where the 
air literally drags. The roughness of the platter surface is captured in $C_d$.

\subsection{Power}
In its mechanical form, 
power is the rate of doing work ($W$) or expending energy.
\begin{equation}
\mathcal P = \dfrac{dW}{dt}  \, . \label{eqn:mech-pwr}
\end{equation}
For an HDD, however, it's probably more convenient to measure the 
energy loss due to aerodynamic friction in terms of the current drawn ($I$) 
at voltage ($V$):
\begin{equation}
\mathcal P = I V = \dfrac{dq}{dt} V \, .  \label{eqn:elec-pwr}
\end{equation}
If we count the charges $q$ as electrons, then $dq V$ is measured in 
electron-volts (eV), 
which is a measure of energy\footnote{KeV in an analog TV.}.
Hence, \eqref{eqn:mech-pwr} and \eqref{eqn:elec-pwr} are 
dimensionally equivalent.

Moreover, the work performed in \eqref{eqn:mech-pwr} is by virtue of the 
friction force in \eqref{eqn:drag} acting along an elemental (tangential) arc length $ds$,
i.e., $dW = F_{\rho} \, ds$. Hence, the power
\begin{equation}
\mathcal P = F_{\rho} \, \dfrac{ds}{dt} = F_{\rho} \, v  \, , \label{eqn:bernoulli-pwr}
\end{equation}
can be expressed directly in terms of the drag force and the tangential velocity.

\begin{figure}[!ht]
\centering
\includegraphics[scale = 0.45]{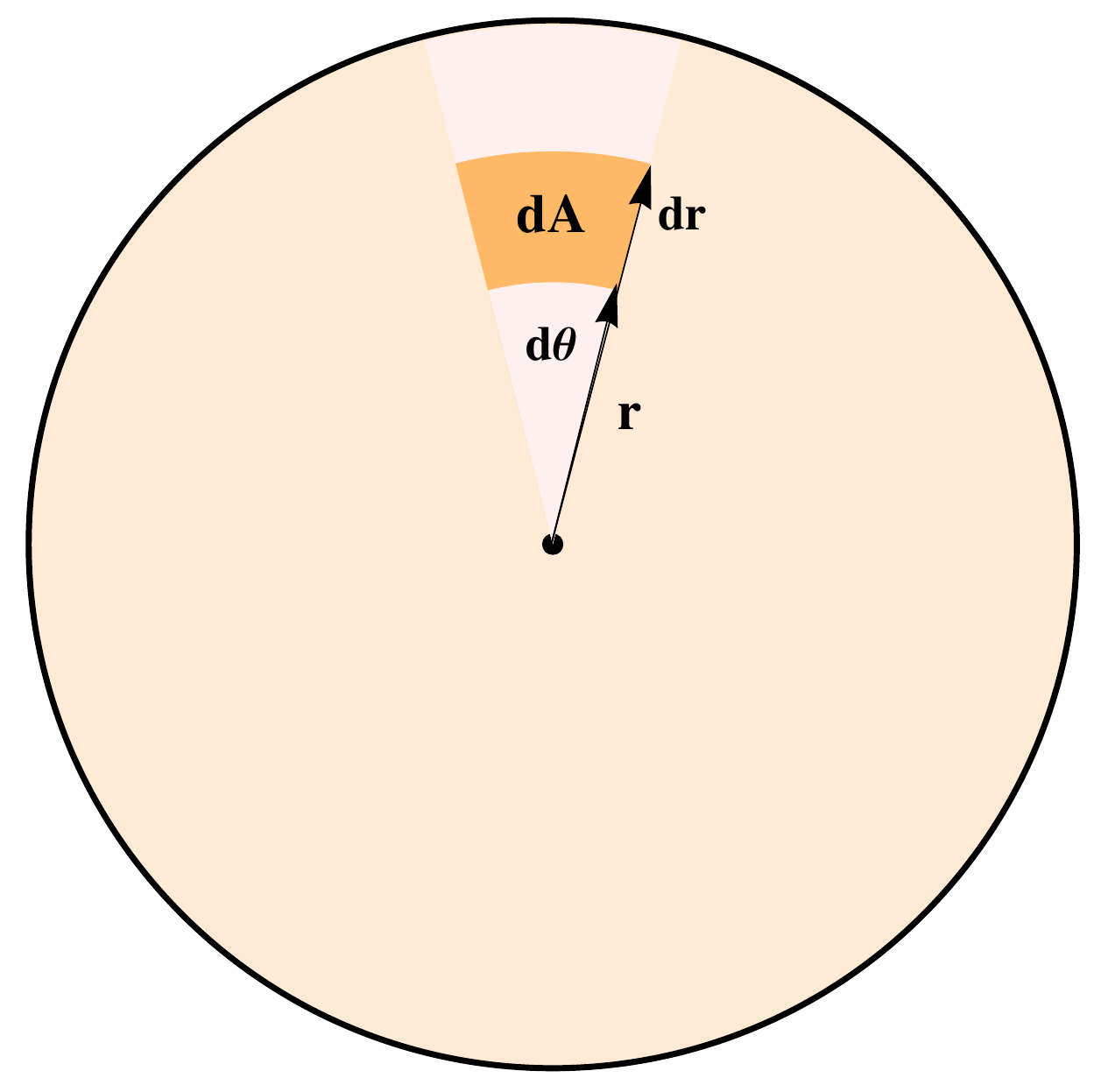} 
\caption{Differential area element $dA$ defined on a disk platter}
\label{fig:disksector}
\end{figure}

However, each elemental area ($dA$) on the platter moves with a different tangential speed 
and therefore experiences a different force.
Thus, we have to integrate all possible patches over the whole platter. 
Applying \eqref{eqn:drag}, the power integral can be written as:
\begin{equation}
\mathcal P \propto \int \rho \, v^3 \, dA   \, , \label{eqn:totpwr} 
\end{equation}
ignoring any proportionality constant.

From Fig.~\ref{fig:disksector} the sectorial area on the platter can be written as the
product of the arc delta ($ds$) and radial delta ($dr$):
\begin{equation}
dA = ds . dr = (r d\theta) dr  \, .  \label{eqn:delarea} 
\end{equation}
Similarly, the tangential velocity vector $\mathbf v$ is related to the 
(axial) angular velocity $\mathbf \Omega$ by $\mathbf v = \mathbf \Omega \times \mathbf r$ but,
since all these vectors are orthogonal on the disk we can use the scalar form
\begin{equation}
v = |\mathbf\Omega| r \,  . \label{eqn:sangvel} 
\end{equation}
Substituting \eqref{eqn:delarea} and \eqref{eqn:sangvel} into \eqref{eqn:totpwr} produces 
the two-dimensional integral
\begin{equation*}
\mathcal P \propto \rho \int_0^R dr \; \int_0^{2\pi} r \, d\theta \; r^3 \Omega^3 
  \equiv \rho \, \Omega^3 \int_0^R dr \, r^4 \; \int_0^{2\pi} d\theta  \, ,
\end{equation*}
which yields
\begin{equation}
\mathcal P \propto \dfrac{2\pi}{5} \rho \, \Omega^3 \, R^5  \label{eqn:totpwr2} \, .
\end{equation}
With $D=2 R$, \eqref{eqn:totpwr2} is identical to \eqref{eqn:emp-model} up to constants of proportionality.

\section{Summary}
The theoretical contributions to \eqref{eqn:emp-model} can be summarize in the following simple steps:
\begin{enumerate}
\item Factor the power $\mathcal P$ as $\Omega^3 R^3 \times R^2$
\item $R^2$ comes from the total area of the platter
\item $\Omega R$ is the angular speed $v$ of air at any point on the spinning platter
\item Why the cube: $v^3 \equiv \Omega^3 R^3$ ?
\item A factor $v^2$ comes from the Bernoulli pressure ($F/A$) in \eqref{eqn:drag}
\item Another $v$ factor comes from the definition of power $\mathcal P = v \times F/A$ in \eqref{eqn:bernoulli-pwr}
\end{enumerate}

A factor of $2$ should also included since the drag will occur on both the 
upper and lower side of the platter, but we are not keeping track of constants here.
There could also be inter-platter turbulence but the 
Reynold's number is likely already substantial due to head movement.

It has also been shown recently that electrical power consumption is
proportional to the third power of the logical block number~\cite{ukcam}. This
is likely related to $\Omega^3$ via \eqref{eqn:sangvel}.

The preceding discussion concerning HDD aerodynamic power consumption will become moot as SSDs 
become increasingly cost-effective and reliable~\cite{ssdtom,ssdvhdd,anand}.

\end{document}